\documentclass[runningheads,pdflatex]{llncs}
\usepackage[pdftex]{graphicx}
\usepackage{comment}

\usepackage{amssymb,amsmath}
\usepackage{cancel}

\usepackage{url}

\begin{document}

%\pagenumbering{arabic}
%\pagestyle{plain}

\setlength\abovecaptionskip{0pt}
\setlength\textfloatsep{10pt}

\title{Conflict-free Collaborative Set Sharing\\
for Distributed Systems}
%titlerunning, author, ... for llncs
\titlerunning{Conflict-free Collaborative Set Sharing}

\author{Masato Takeichi}
\authorrunning{M. Takeichi}
%\institute{University of Tokyo\\
\institute{University of Tokyo\\
\email{takeichi@acm.org}
}
%\date{\today}
%\date{November 19, 2021}

\maketitle

% llncs abstract
\begin{abstract}
Collaborative Data Sharing is widely noticed to be
essential for distributed systems.
Among several proposed strategies,
{\em conflict-free} techniques are considered
useful for serverless concurrent systems.

They aim at making shared data be consistent
between peers in such a way that their local data
do not become equal at once, but they arrive at
the same data eventually when no updates occur
in any peer.

Although the {\em Conflict-free Replicated Data Type}
(CRDT)
approach
could be used in data sharing as well,
it puts restrictions on
available operations so as to concurrent updates
never cause conflicts. Even for sets, popular operations
such as
insertion and deletion
are not freely used, for example.

We propose a novel scheme for {\em Conflict-free
Collaborative Set Sharing} that allows both
insertion and deletion operations.
It will provide a new synchronization method for
data sharing and gives a fresh insight into
designing conflict-free replicated
data types.
We might consider that this becomes
a substitute for CRDTs.
\end{abstract}

\section{Introduction}
\label{sec:Introduction}
In the server-client system, clients can easily share
data on the server.
This is a traditional style of data sharing.
However, in distributed systems where each site, or
peer, has its own exclusive property of the contents
and the policy,
data sharing between peers for
collaborative work raises several substantial
problems.

We have been discussing
``What should be shared'' in collaborative data
sharing,
but not so much talking
about ``How should be shared''.

Concerning the ``what'',
a seminal work on
{\em Collaborative Data Sharing}
~\cite{Ives2005,Karvounarakis2013}
brought several issues upon specification of
data to be shared.
An approach based on
the view-updating technique
with {\em Bidirectional Transformation}
\cite{Lenses,Bohannon:08,Hidaka:10,HuMT08,HSST11}
has been proved promissing.
Among others and
aside from data transformation,
we observe different approaches on the ``how'':
The {\em Dejima} 1.0 implementation
\cite{dejima-theory:2019,SFDI-Asano2020}
uses the PostgreSQL
synchronization mechanism with {\em strong consistency}
between distributed data.
And the {\em BCDS Agent}~\cite{Takeichi21JSSST}
implementation applies
a ``happy-go-lucky'' technique for
{\em eventual consistency}
based on the property of {\em Bidirectional Function
Composition}~\cite{Takeichi09IFL},
but we face some difficulty to foresee
the final result because of its global nature.

In collaborative data sharing, each peer has its
local data and it provides some of them to other
peers and receives some data from peers
for reflecting updates on these shared data.

From the other side of data sharing,
peers may start with the common data as shared
and locate its replicas as their own local data.
And then each peer performs local operations on the
replica and does any way to synchronize it with other
peers.
In this course, how to share data between peers
is same as
how to synchronize distributed replicas to be the same.
Thus, they are almost equivalent except
original intentions.
And our problem to be solved
is how to synchronize distributed replicas in
serverless
distributed systems.

We have various kinds of
{\em Conflict-free Replicated Data Types}
(CRDTs)~\cite{Shapiro2011CRDTs}.
The CRDT approach restricts
available operations acted on replicated data;
the {\em Grow-Only-Set} (G-Set) CRDT allows only
the insertion operation on the set data, for example.

We can observe another kind of collaborative systems
in daily life, i.e.,
Realtime Collaborative Document Editors.
Most systems are implemented with the use of the
{\em Operational Transformation} (OT)
technique~\cite{Ellis1989,Sun1998}.
The OT for text editing deals with a string as the
replica and the operations on the replica should be
aware of the position in the string.
The synchronization process is known to be
rather complex and error-prone.
And most of these systems use the server for
synchronization.
So, it is not appropriate for us to add emphasis
on OT for data sharing in general, while we
come across its basic idea in our scheme.

In this paper, we will explore a novel scheme
for set data sharing between distributed replicas.
The set data is the base of various data types
and it is ubiquetous in that it spreads
over many applications and it can be extended
in many ways.

Our {\em Conflict-free Collaborative Set Sharing}
(CCSS)
allows general set operations, i.e., insertion
and deletion of an element, and avoids any
conflicts between concurrent operations to
realizes eventual consistency.
This is the most distinguished feature of our CCSS
compared with CRDTs.

\section{Whereabouts of Conflicts in Distributed Systems}
\label{sec:Whereabouts}
Consider a data sharing example of distributed systems:
Peers $P$ and $Q$ have their local data
$D_P$ and $D_Q$ to be appropriately {\em synchronized}.
That is, $P$ and $Q$ have replicas $D_P$ and $D_Q$
respectively as instances of the same set.
$P$ and $Q$ update $D_P$ and $D_Q$ respectively
whether or not the network connection is alive,
and they try to synchronize them during the
connection is alive.

Each peer inserts element $x$ into its local data
(written as $\cup \{x\}$) and deletes element
from it (written as $\setminus \{x\}$),
and sends the operations thus performed to the
partner peer. This is the {\em client} function
of the peer.
The peer as the {\em server} receives remote
operations from the partner peer and puts them on
the local data so that it becomes same as that of
the partner peer.

What happens in the events?
\vspace{-0.2cm}
\begin{enumerate}
  \item Start with $D_P=D_Q=\{1,2\}$.
  \item Network connection fails.
  \item $P$ does $\cup \{3\}$ and then $\setminus \{3\}$.
  \item $Q$ does $\setminus \{2\}$ and then $\cup \{3\}$.
  \item Connection is restored.
\end{enumerate}
How are $D_P$ and $D_Q$ synchronized?
And what is the result after Step 5?
Is $D_P=\{1,2\} {\rm ~or~}\{1,3\}$?
Is $D_Q=\{1,3\} {\rm ~or~}\{1,2,3\}$?

It would be reasonable to answer this question with
expectation as
``It should be $D_P=D_Q=\{1,3\}$.''

This small example may remind us of conflict resolutions
for data sharing in distributed systems.

\subsection{How CRDT Solves the Problem}
\label{sec:CRDT}
The Conflict-free Replicated Data Type (CRDT)
approach~\cite{Shapiro2011CRDTs}
follows ``When in Rome do as the Romans do''.
That is, the local data $D$ of type CRDT is defined
by restricting operations on $D$
so that it never becomes inconsistent with others
upon updates.

There are two types of CRDT approaches:
operation-based CRDTs and state-based CRDTs.
It is known that the above two are equivalent.
Since we will propose a scheme based on operations,
we give here an overview of operation-based CRDTs.

Operation-based (op-based) CRDTs ~\cite{Preguica2018crdts}
places
data type operations into messages,
which are sent to all replicas in
order.
The peers apply received operations to their replica
so that they all arrive at the same state,
even if they receive concurrent messages in different
orders.

The {\em Grow-Only-Set} (G-Set) allows only the
insertion operation $\cup \{x\}$\footnote{
The operation is applied from the left of the operand
$D$, i.e., $D \cup \{x\}$ represents the set data
obtained by inserting $x$ into $D$. Thus, all
the operations in this paper are postfixed after
the operand data.
}.

The local data of $P$ and $Q$ were synchronized
before connection failure and they have been
modified with local operations
$\langle \cup \{p_1\},\cup \{p_2\}, \cdots,
\cup \{p_m\}  \rangle$
and
$\langle \cup \{q_1\},\cup \{q_2\}, \cdots,
\cup \{q_n\}  \rangle$
respectively during the connection failure period.

After the connection is established again,
$P$ and $Q$ send their local operations to each other
for synchronization.
Then, $P$ as the server applies the remote operations
received from $Q$ to the current local data
$D_P \cup \{p_1\} \cup \{p_2\} \cdots \cup \{p_m\}$
to obtain
$D_P \cup \{p_1\} \cup \{p_2\} \cdots \cup \{p_m\}
\cup \{q_1\}\cup \{q_2\} \cdots \cup \{q_n\}$.
Similarly $Q$ has now the new local data
$D_Q \cup \{q_1\} \cup \{q_2\} \cdots \cup \{q_n\}
\cup \{p_1\}\cup \{p_2\} \cdots \cup \{p_m\}$.
It is easy to show that these are same provided
that $D_P=D_Q$ before the connection failure.
This is because
the commutative property
$D\cup \{x\}\cup \{y\}=D\cup \{y\}\cup \{x\}$
holds for any $x$ and $y$.

As illustrated above in the G-Set CRDT, CRDTs solve
consistency problems by only allowing {\em monotonic}
updating operations; any operation must make the
structure larger.

To define a Set CRDT with insertion and deletion,
we have to do something for deletion since deletion
breaks monotonicity.
A simple idea called {\em Two-Phase-Set} (2P-Set) CRDT
is to use a pair $(A,R)$ of
two G-Sets for the local data $D$: $A$ for inserted
(added) elements and $R$ for deleted (removed) elements.
Deletion never actually removes elements,
but does mark them as deleted and keep them in the
second G-Set $R$ of ``tombstones''.
When an element $x$ in the local data $D=(A,R)$
is ``deleted'', $A$ remains unchanged, i.e., $A$ does not
become $A\setminus \{x\}$, while $R$ grows to
$R\cup \{x\}$. Thus, the deletion operation is
also monotonic and $D$ does not shrink.

Then, how can we answer the question:
``What is the actual elements of the local data $D$?''
In this CRDT, we should answer that
``It is $A \setminus R$''.
That is, all the elements of
$A$ not included in $R$, are the actual existent elements.

We cannot effectively
insert elements again into the
2P-Set after they have been deleted sometime before,
since $A \setminus R$ always excludes elements
ever deleted from $A$.

As an oft-cited example of a shopping basket
in an online shop:
\vspace{-0.2cm}
\begin{itemize}
  \item The G-Set cannot be used, for we cannot remove
  items once added to the cart.
  \item The 2P-Set cannot be practical, for we might
  finally want to add items that were once added to
  the cart but deleted sometime before.
\end{itemize}

For the example above in this Section,
the 2P-Set CRDT may cause difficulties.
If we use 2P-Sets for $D_P$ and $D_Q$,
starting from $D_P=D_Q=\{1,2\}$,
after
$P$ does
$\cup \{3\}$ and then $\setminus \{3\}$
and
$Q$ does
$\setminus \{2\}$ and then $\cup \{3\}$,
we see that $3$ is never included actually
in $D_P$ and $D_Q$ since $P$'s operation
$\setminus \{3\}$
rejects actual $3$ whether it precedes
or follows $Q$'s operation $\cup \{3\}$.

From these observations,
we note that
we should develop another schema for sharing set data
with useful operations available,
insertion and deletion, under our
usual interpretation.

\section{Conflict-free
Collaborative Set Sharing}
\label{sec:CCSS}
The basic idea behind our
{\em Conflict-free Collaborative Set Sharing} (CCSS)
is to take particular
note on the fact that our object data is {\em mutable}
and operations on that data should be closely
related to the
current state (value) of the data.

Usual mathematical definitions of operations
$\odot x$ on set $D$
refers to the operator $\odot$ that maps as
$(D,x) \mapsto D'$ independently of $D$ with $x$.
But we define here $\odot x$ itself by
making full use of the
relationship between $D$ and $x$ like $x \in D$,
for example.

Consider another small example using usual
set operators:
\vspace{-0.2cm}
\begin{itemize}
  \item Assume that $P$ and $Q$ have local data $D_P$
  and $D_Q$ which have been synchronized as
  $D_P=D_Q=\{1,2\}$.
  \item And then, $P$ wants to ``insert $2$''
  into $D_P$ by a postfix operation $\cup \{2\}$ and
  $Q$ tries to ``delete $2$'' from $D_Q$
  by operation $\setminus \{2\}$.
  \item When both operations finish, $D_P$ remains
  unchanged while $D_Q$ has been changed into $\{1\}$.
\end{itemize}
How can we make $D_P$ and $D_Q$ be synchronized,
i.e., are made the same?
We have no clue for synchronization
unless something is given other than the current data.
If we are given the operations
$\cup \{2\}$ on $D_P$ and $\setminus \{2\}$ on $D_Q$,
we can use them for understanding the intenstions.

Since $D_P=D_Q$ when $P$ and $Q$ began to perform
concurrent operations, we should observe
that $\cup \{2\}$ and $\setminus \{2\}$
cause a {\em conflict} of effective update:
Which should be taken for obtaining consistent
$D_P$ and $D_Q$?

To answer this question, consider the reason
why $P$ wanted to ``insert $2$'' into $D_P$.
Supposedly, $P$ wanted to share $2$ with $Q$
by means of and at the time of the next synchronization.
As it were, what happened if $P$ had checked before
taking the operation whether ``$2 \in D_P$''?
If $P$ noticed that $2$ was already in $D_P$,
$P$ had nothing to do for that
purpose because
the intended state had already been there.

If it were, no conflict occurs!

This is what leads us to the idea of using
{\em effectful} set operations for
our Conflict-free Collaborative Set Sharing.

\subsection{Effectful Set Operations}
We assume that our mutable data $D$ is a set of
elements $x$ of any type,
and operations $\oplus x$ and $\ominus x$
on $D$ change the value $D$ into $D \oplus x$
and $D \ominus x$, respectively.
Of course, we can read this mathematically
as $D$ is mapped to
$D \odot x$, i.e., $(D, x) \mapsto D \odot x$ by
$\odot x$.
We are using a generic operator symbol $\odot$
for representing $\oplus$ or $\ominus$.

We call insertion operation $\oplus x$ {\em effectful}
if it gives $D\oplus x \ne D$.
That is,
the effectful operation $\oplus x$ is defined
and it can be applied to $D$ only if $x \not \in D$.
Similarly the effectful $\ominus x$ is defined
and it can be applied to $D$ only if $x \in D$, and then
$D \ominus x \ne D$.

In short, the effectful operations always change $D$
when they are defined and applied to $D$,
while usual set operation $D \cup \{x\}$ and
$D \setminus \{x\}$ do not always.

For convenience, we introduce a postfix identity
operation ``$!$'' which does not change the value, i.e.,
$D!=D$ for any $D$.
In fact the operation ``$!$'' is not effectful
according to the
above intuitive meaning, but we will use this for the
``do nothing'' operation.

\subsubsection*{Properties of effectful set operations}
For the operations $\oplus x$ and $\ominus x$,
\vspace{-0.2cm}
\begin{itemize}
  \item $D \oplus x \oplus x$ and $D \ominus x \ominus x$
  {\em never appear} since the second occurences
  of $\oplus x$ and
  $\ominus x$ after the first same operation
  are {\em not defined}.
  Thus, the validity of using the effectful
  operations depends on the context.
  \item $D \ominus x \oplus x =D$ and
  $D \oplus x \ominus x$ hold as long as they are valid,
  i.e, $x \in D$ in the first case, and
  $x \not \in D$ in the second case.
  We can read this as effectful $\oplus x$ and $\ominus x$
  cancel each other.
  \item For $x \ne y$, hold
  \vspace{-0.2cm}
  \begin{equation*}
    \begin{split}
      D \oplus x \oplus y &= D \oplus y \oplus x\\
      D \oplus x \ominus y &= D \ominus y \oplus x\\
      D \ominus x \ominus y &= D \ominus y \ominus x
    \end{split}
  \end{equation*}
  This means that $(\oplus x,\oplus y)$,
  $(\oplus x,\ominus y)$ and $(\ominus x,\ominus x)$
  are {\em commutative}.
\end{itemize}

\subsection{Normalization of Operation Sequence}
We write a sequence of operations as
$\langle \odot x_1, \odot x_2, \cdots \rangle$
where each $\odot$ may not be the same; so it represents
$\langle \oplus x_1, \oplus x_2, \cdots \rangle$,
$\langle \oplus x_1, \ominus x_2, \cdots \rangle$,
..., etc.

From the above canceling rule $D \oplus x \ominus x$
and the commutativity rule of
$\oplus x$ and $\ominus x$,
if $\oplus x$ and $\ominus x$ appear in this order
with no $\odot x$ in-between,
\begin{equation*}
  D\odot \cdots \oplus x \cdots \ominus x \cdots
  =
  D \odot \cdots ! \cdots ! \cdots
\end{equation*}
holds.
In fact, the occurences of $\oplus x$ and $\ominus x$
may be removed from the sequence.
But we use the identity operation ``$!$'' to
fill the positions
to keep the length of the sequence.

Same for the case that $\ominus x$ and $\oplus x$
appear in this order.

Thus, we can {\em normalize} the operation sequence
into one that does not contain cancel-able pairs of
$\oplus x$ and $\ominus x$, and they are
replaced with ``$!$''.
This does not cause any effect on the data.

And from now on, we assume that the operation
sequence has been normalized.

As a consequence of normilization,
no duplicate appears in the normalized sequence
$\langle \odot x_1, \odot x_2, \cdots \rangle$.
This is because that if ever there were pairs of
operations satisfying $x_i=x_j$, i.e., the same operation
appear at different positions,
they must be $(\oplus x_i, \ominus x_j)$
or
$(\ominus x_i, \oplus x_j)$,
since neither $(\oplus x_i, \oplus x_j)$
nor $(\ominus x_i, \ominus x_j)$ appears from the
definition of effectful operations.
But this contradicts the assumption that the operation
sequence has been normalized.

\subsection{Synchronization of Effectful Operations}
Assume that $P$ and $Q$ share data $D$ by locating its
replicas $D_P$ and $D_Q$ as their local data. And they
independently and concurrently perform local operations
$\odot p$ and $\odot q$ respectively on their replicas.

Also assume that $P$ as the client has performed local
operations $\odot p_1$, $\odot p_2$, …, $\odot p_m$
on local data $D_P$ to get the current data
$D_P \odot p_1 \odot p_2\odot \cdots \odot p_m$.
Similarly $Q$ has got
$D_Q \odot q_1 \odot q_2\odot \cdots \odot q_n$
by operations
$\odot q_1$, $\odot q_2$, …, $\odot q_n$.

Then, what should be done for synchronizing $P$'s replica
and $Q$'s replica so that they contain the same data?

$P$ as the server receives remote operations
$\odot q_1$, $\odot q_2$, …, $\odot q_n$ from $Q$
to make the local data reflect these remote operations.
A simple-minded way to do this might be applying remote
operations to the current data as
\begin{equation*}
D_P \odot p_1 \odot p_2\odot \cdots \odot p_m
\odot q_1 \odot q_2\odot \cdots \odot q_n.
\end{equation*}
However, this sometimes fails because the effectful
operation $\odot q_j$ ($j=1,2,\cdots, n$) is not
always valid in
this expression.
Recall that the effectful $\oplus x$ can be applied to
$D$ only if $x \not \in D$ and
$\ominus x$ can be applied to $D$ only if $x \in D$,
but $\odot q_j$ may violate these side conditions when
it is applied to $P$'s replica,
while $\odot q_j$ is valid in $Q$'s replica.

Therefore we need to transform each remote operation
$q_j$ into an effectful $q_j'$ that reflects the
effect of $\odot q_j$ on the current data.

\subsubsection*{Confluence of updates by synchronization}
Given normalized operation sequences
$\boldsymbol
{ps} =\langle \odot p_1, \odot p_2, \cdots , \odot p_m\rangle$
and
$\boldsymbol{qs}=\langle \odot q_1, \odot q_2, \cdots ,\odot q_n\rangle$,
$P$ as the server calculates
$\boldsymbol{qs}'=\langle \odot q_1', \odot q_2', \cdots ,\odot q_n'\rangle$
according to the following rule:
For each $j=1,2, \cdots, n$, if operation $q_j$ appears
in $\boldsymbol{ps}$,
then set $q_j':=!$
else set $q_j':=q_j$.

Using this $\boldsymbol{qs'}$, the current data in $P$
is updated into
\begin{equation}
  \label{D_Ppsqs'}
  D_P \odot p_1 \odot p_2\odot \cdots \odot p_m
  \odot q_1' \odot q_2' \odot \cdots \odot q_n'.
\end{equation}
This expression does not violate the validity of effectful
operations.

It is almost the same in $Q$. We can apply the same
algorithm by exchanging $\boldsymbol{ps}$ and
$\boldsymbol{qs}$ and calculating $\boldsymbol{ps'}$.
The current data in $Q$ is now updated into
\begin{equation}
  \label{D_Qqsps'}
D_Q \odot q_1 \odot q_2\odot \cdots \odot q_n
\odot p_1' \odot p_2' \odot \cdots \odot p_m'.
\end{equation}

Apart from procedural operations, we can do
another transformation in $P$ as if its local data
were not yet updated by $\boldsymbol{ps}$.
The above algorithm for $P$ can be rewritten for
obtaining $\boldsymbol{ps'}$ as:
For each $j=1,2, \cdots, n$, if operation $q_j$ appears
at some position, say $k$ in $\boldsymbol{ps}$,
then set $p_k':=!$ else set
$p_k':=p_k$.

As long as the final value is concerned,
we have
\begin{equation}
  \label{D_Pps'qs}
  D_P \odot p_1' \odot p_2' \odot \cdots \odot p_m'
  \odot q_1 \odot q_2 \odot \cdots \odot q_n.
\end{equation}
by first applying operations $\boldsymbol{ps'}$
and then applying operations $\boldsymbol{qs}$.

From the commutative property of the effectful
operations, and the fact that the operation sequences
$\boldsymbol{ps}$ and $\boldsymbol{qs}$ have been
normalized, we conclude that
\vspace{-0.2cm}
\begin{itemize}
  \item The data value \eqref{D_Pps'qs}
  is equivalent to \eqref{D_Ppsqs'}.
  This is because that by exchanging the role
  of $q_j'$ in \eqref{D_Ppsqs'} and $p_k'$
  \eqref{D_Pps'qs},
  we can get the same data value.
  \item The data value \eqref{D_Pps'qs} is equivalent to
  \eqref{D_Qqsps'} provided $D_P=D_Q=D$.
  This is because that by moving $q_j$ in
  \eqref{D_Pps'qs}
  to the left before $p_i'$ we have \eqref{D_Qqsps'}.
\end{itemize}

\begin{figure}[htb]
 \centering
  \includegraphics
  [width=\linewidth]{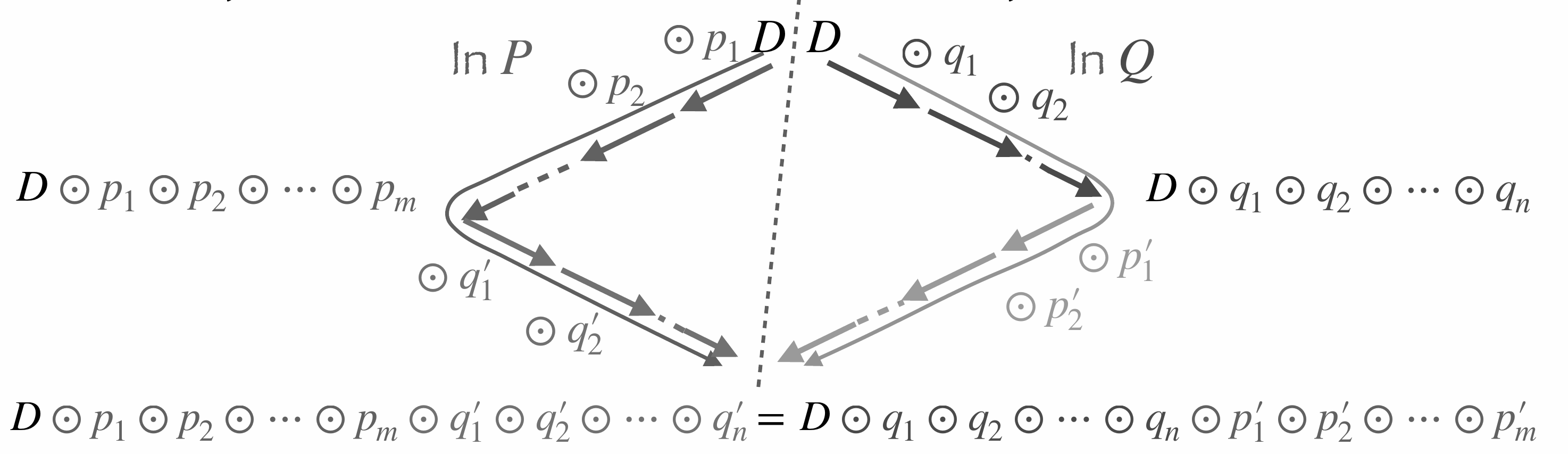}
  \vspace*{0pt}
  \caption{Confluence of Updates by Synchronization}
  \label{fig:Confluence}
\end{figure}

Hence, the replicas after independent
synchronization by $P$
and $Q$ have the same data value.
That is, our transformation assures the confluence
property of the effectful set
operations (Fig.\ref{fig:Confluence}).

Recall the small example in Section \ref{sec:Whereabouts}:
$P$ and $Q$ start with $D_P=D_Q=\{1,2\}$, and
$P$ does $\cup \{3\}$ and then $\setminus \{3\}$,
and $Q$ does $\setminus \{2\}$ and then $\cup \{3\}$.
We see that operations $\cup \{x\}$ and
$\setminus \{x\}$ here are in fact effectful.
So, $P$ performs operations
$\boldsymbol{ps}=\langle \oplus 3, \ominus 3 \rangle$
to have $D_P'=\{1,2\}\oplus 3\ominus 3=\{1,2\}$,
and $Q$ performs concurrentry
$\boldsymbol{qs}=\langle \ominus 2, \oplus 3\rangle$
to have $D_Q'=\{1,2\}\ominus 2\oplus 3=\{1,3\}$.
As our first step to do is to normalize the operation
sequences:
$\boldsymbol{ps}$ becomes $\langle !, ! \rangle$ and
$\boldsymbol{qs}$ remains as it is.
The above synchronization procedure derives
$\boldsymbol{qs'}=\boldsymbol{qs}$ since there is no
elements in $\boldsymbol{ps}$ that is equal to $q_j$.
Hence, $P$'s local data becomes
$D_P'\ominus 2\oplus 3=\{1,2\}\ominus 2\oplus 3=\{1,3\}$.
$Q$ produces $\boldsymbol{ps'}=\langle !, ! \rangle$
from $\boldsymbol{ps}$, and gives local data
$D_Q'~!~!=\{1,3\}~!~!=\{1,3\}$.
Thus our synchronization gives a confulence $\{1,3\}$
as expected.

\subsubsection*{A Digression}
\label{sec:Quiz}
Given lists of integers $xs=[x_1, x_2, ...,x_m]$ and
$ys=[y_1, y_2, ...,y_n]$ each has no duplicate elements
in itself, but with possible duplicates between
$xs$ and $ys$.
Then, how can we calculate the sum of different
integers in concatenated list $xs+\!+ys$?
For example, $xs=[3,1,4,5,9,2]$ and $ys=[8,2,7,6,1]$
have no duplicates in themselves,
but $1$ and $2$ appear in $xs+\!+ys$.

We may write code\footnote{
Haskellers may solve this quiz by
$sum \cdot nub$.
Haskell's standard library provides function
$nub$ for removing duplicates in a list.
But this does not help us here to understand
our synchronization.
}:
\vspace {-0.2cm}
\begin{itemize}
  \item Compute first $ys'=[y_1', y_2',...y_n']$ by
  replacing $y_j$ with $0$ if it appears in $xs$
  or keeping it in $ys'$
  for $j=1,2,\cdots$.
  \item And then compute $sum (xs+\!+ys')$.
\end{itemize}
For the example above,
$sum ([3,1,4,5,9,2]+\!+[8,0,7,6,0])$ gives the answer.
Note that by computing $xs'$ from $xs$ and $ys$,
$sum ([3,0,4,5,9,0]+\!+[8,2,7,6,1])$
gives the same result.

This algorithm helps us
to understand our procedure
for our conflict-free synchronization of set operations.

\subsection{Eventual Consistency over Distributed Peers}
Note that update synchronization in each peer
does not necessarily processed at the same time,
rather each peer does it when convenient.
We can see that independent update synchronization
eventually arrives at the same data after no more
local operations are performed in both
under the conditions:
\vspace{-0.2cm}
\begin{itemize}
  \item $D_P$ and $D_Q$ have been synchronized
  at least once, and
  \item All the local operations $\boldsymbol{ps}$
  in $P$ and
  $\boldsymbol{qs}$
  in $Q$ performed since the last synchronization
  are sent to and received from each other
  with the order kept and no element lost.
\end{itemize}

The above observation comes from the fact that our procedure for synchronization to compute
\begin{equation*}
    D_P \odot p_1 \odot p_2\odot \cdots \odot p_m
    \odot q_1' \odot q_2' \odot \cdots \odot q_n'.
\end{equation*}
from $\boldsymbol{ps}$ and $\boldsymbol{qs}$
can be divided into segments in any ways such as
\begin{equation*}
    D_P \odot p_1 \odot q_1'
    \odot p_2\odot \cdots \odot p_m
    \odot q_2' \odot \cdots \odot q_n'.
\end{equation*}

\subsubsection*{Managament of Local and
Remote Operations}
To synchronize the local data, or replicas of
independent peers, we have to know
the shared data which are synchronized last time.
These are the roots to which local operations
performed and then followed by remote operationsfor synchronization.

Since $P$ and $Q$ concurrently run,
they do not always perform
updates at the same time.
So, they need to keep operations since the last
synchronization until the next.
Then, when and how can we shorten the operation sequence?

\begin{figure}[htb]
 \centering
  \includegraphics
  [width=\linewidth]{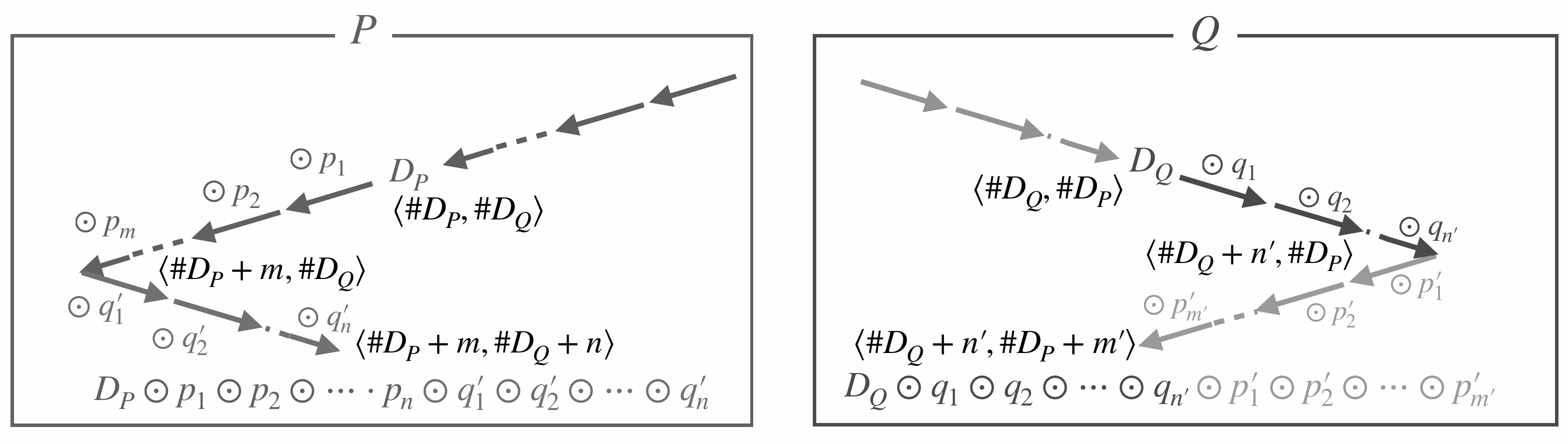}
  \vspace*{0pt}
  \caption{Revision Numbers with Local
  and Remote Operations}
  \label{fig:RevisionNumber}
\end{figure}

The revision number $\#D$ of the local data $D$
helps us to recognize the state of synchronization.
It is incremented every time a local operation $p_i$
is performed
including transformed remote operations $q_j'$.
By maintaining the pair of local and revision
numbers $\langle \#D_P, \#D_Q \rangle$ of $P$
and sending this with operations to the partner $Q$,
we can recognize which part of local operations are no
more needed (Fig.\ref{fig:RevisionNumber})

\subsubsection*{Synchronization of Distributed Peers}
So far, we have seen solely how synchronization works
between $P$ and $Q$.

If $P$ has another connection to peer $R$,
$P$ needs to synchronize $P$'s local data
and $R$'s local data as $P$ has done with $Q$.

Consider $P$ begins to synchronize its current
local data just after the synchronization with $Q$.
The synchronization with $Q$ began with
revision numbers
$\langle \#D_P, \#D_Q \rangle$ and local data
$D_P \odot \boldsymbol{ps}$ and produced
the local data
$D_P \odot \boldsymbol{ps}\odot \boldsymbol{qs'} $,
where $\odot \boldsymbol{ps}$ represents applying
operations of $\boldsymbol{ps}$.

The last synchronized data of $P$ with $R$ is not
necessarily the same as with $Q$. And operation
sequence $\boldsymbol{qs'}$ performed in the
synchronization with $Q$ has been sent to $R$
as the propagation of $Q$'s updates to other peers.

Let $D_P'$ and $D_R$ has been synchronized, and
$\boldsymbol{ps*}$ is the operations performed in $P$
satisfying
$D_P' \odot \boldsymbol{ps*} =
D_P \odot \boldsymbol{ps}\odot \boldsymbol{qs'} $.

\begin{figure}[htb]
 \centering
  \includegraphics
  [width=0.75\linewidth]{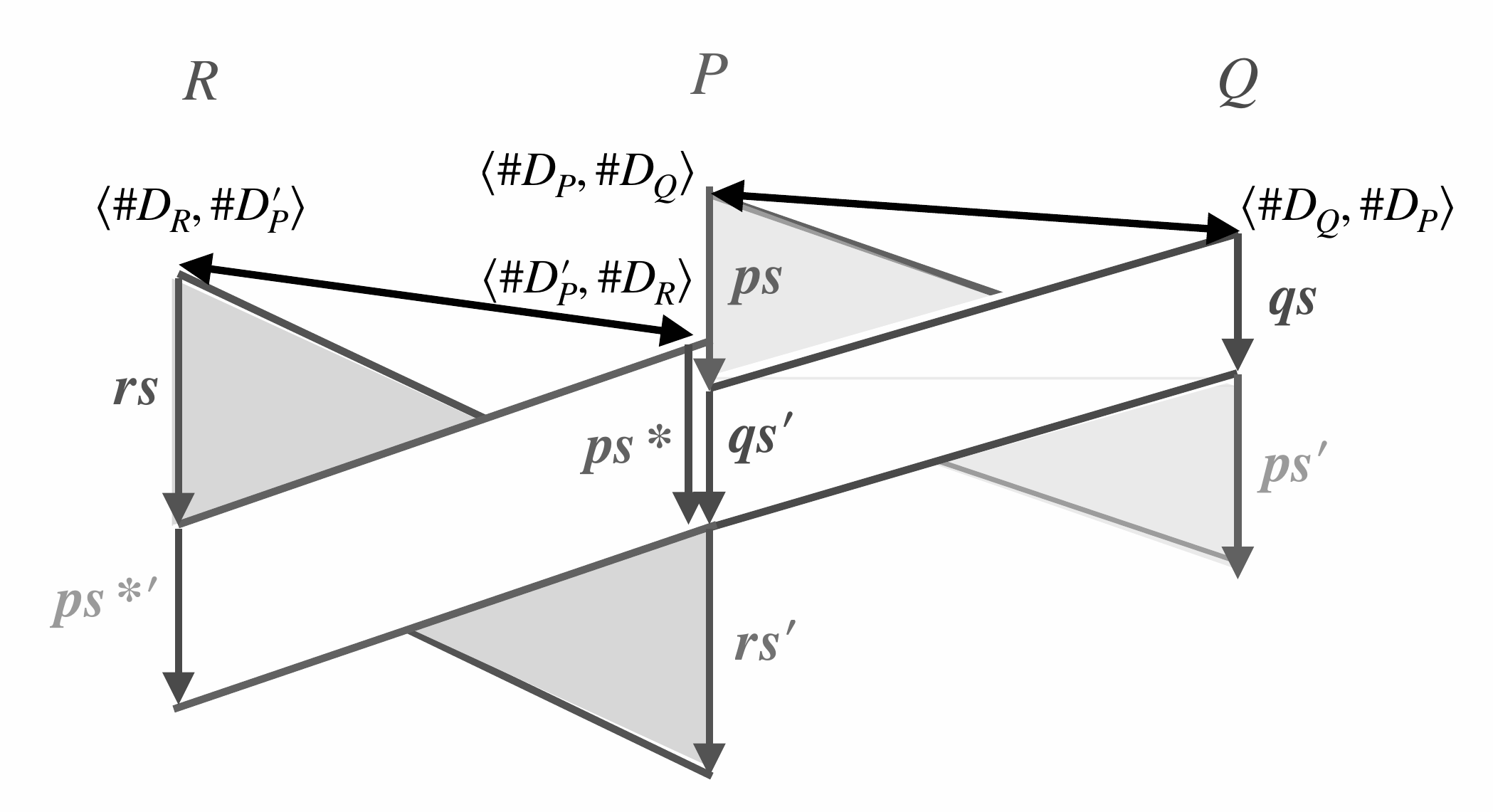}
  \vspace*{0pt}
  \caption{Synchronization of Distributed Peers}
  \label{fig:Synchronization}
\end{figure}

Then, the synchronization process starts from
$D_P'$ and local operations $\boldsymbol{ps*}$
already performed on $D_P'$ and remote operations
$\boldsymbol{rs}$ sent from $R$.
This proceeds the same as $P$ did
for $Q$ with $D_P$, $\boldsymbol{ps}$
and $\boldsymbol{qs}$.
This time, $P$ produces
$D_P' \odot \boldsymbol{ps*}\odot \boldsymbol{rs'}$
with new operations $\boldsymbol{rs'}$ from
$\boldsymbol{rs}$ as shown in
Fig.\ref{fig:Synchronization}.
The operations $\boldsymbol{rs'}$ are also
propagated to other peers including $Q$.

Thus, we conclude that our synchronization scheme
works over distributed peers, and this leads to
the eventual consistency of the whole system.

\section{Remarks}
\label{sec:Remarks}
We can enjoy {\em Conflict-free Collaborative
Set Sharing} (CCSS) simply by
\begin{itemize}
  \item When inserting data,
  first check whether it is not there yet, and
  \item When deleting data, first check whether
  it really is there already.
\end{itemize}
When we use these operations for updates on the
local data, failure of the check tells us that the
operation is {\em invalid} and should not be done,
or rather it tells us that it is not necessary for
our intended updates.

It is very simple to do whatever data with ubiquitous
set semantics. As a matter of course, the oft-used SQL
table is the case.

We can easily extend our CCSS to place transformations
at the gateway of the peer for sending and receiving
operations for controlling shared data.
Data sharing with the {\em Dejima} in BISCUITS
Project is an example~\cite{dejima-theory:2019}.

Also we can extend our CCSS to include mechanisms for
choosing one from grouped data according to preferences,
e.g., when or who inserts the data.
We may call such a strategy {\em Semantic Resolution}.
For example, the LWW Set (Last-Write-Win) CRDT can be
realized by
attaching the logical time stamp as metadata to
the data value with the key for grouping
data. A process is provided for choosing one from
candidates inserted by $\oplus (v,k,t)$,
$\oplus (v',k,t')$, $\cdots$. We can choose $(v,k,t)$
if $t > t'$ by performing local operation
$\ominus (v',k,t')$ for LWW.

Implementation of CCSS peers would be straightforward and an exercise of standard network programming.

\begin{comment}

This work is partially supported by the Japan Society
for the Promotion od Science (JSPS) Grant-in-Aid for
Scientic Research (S) No. 17H06099
``Bidirectional Information
Systems for Collaborative, Updatable, Interoperable,
and Trusted Sharing''.

The author would like to thank to Zhenjiang Hu
conducting this project
and also to the members of the project for
discussion.

\end{comment}

\bibliographystyle{abbrv}
\bibliography{citation}

\begin{comment}
\begin{choshashoukai}
  \choshanopic{Masato Takeichi}
  {Professor Emeritus, University of Tokyo (2011).\
   Emeritus Professor, National Institution for Academic\ Degrees and Quality Enhancement of Higher Education (2018).\
   Council Member (2003-2014) and Vice President (2011-2013) of\
   Science Council of Japan.
   }
\end{choshashoukai}
\end{comment}

\end{document}